\pdfoutput=1
\documentclass[twocolumn]{aastex701}

\usepackage{graphicx}
\usepackage{natbib}
\usepackage{relsize}
\usepackage{textcomp}
\usepackage{upgreek}
\usepackage{amsmath}
\usepackage{amssymb}
\usepackage{calrsfs}
\usepackage{afterpage}
\usepackage{float}
\usepackage{makecell}
\usepackage{hyperref}
\usepackage{comment}
\usepackage{tipa}

\usepackage{booktabs} 
\setcitestyle{notesep={}}

\newcommand{\civ}{C\ensuremath{\,\textsc{iv}}}
\newcommand{\mgii}{Mg\ensuremath{\,\textsc{ii}}}
\newcommand{\hei}{He\ensuremath{\,\textsc{i}}}

\newcommand{\ha}{H\ensuremath{\alpha}}
\newcommand{\hb}{H\ensuremath{\beta}}
\newcommand{\nv}{N\ensuremath{\,\textsc{v}}}
\newcommand{\lya}{Ly\ensuremath{\alpha}}
\newcommand{\lyb}{Ly\ensuremath{\beta}}

\newcommand{\pab}{Pa\ensuremath{\beta}}
\newcommand{\pag}{Pa\ensuremath{\gamma}}
\newcommand{\pad}{Pa\ensuremath{\delta}}
\newcommand{\pae}{Pa\ensuremath{\epsilon}}

\def\lsim{\mathrel{\rlap{\lower 3pt \hbox{$\sim$}} \raise 2.0pt \hbox{$<$}}}
\def\gsim{\mathrel{\rlap{\lower 3pt \hbox{$\sim$}} \raise 2.0pt \hbox{$>$}}}

\def\kms{{\rm km}\,{\rm s}^{-1}}

\def\M1450{{\rm M}_{\rm 1450}}

\begin{document}

\title{\textit{To be lensed or not to be lensed}:\\
on the nature of the alleged high-$z$ lensed quasars J0109--5424 and P170$+$20}



\author[0009-0006-0286-0639]{Aurora Mata-Sanchez}
\affiliation{International Gemini Observatory/NSF NOIRLab, 670 N A`ohoku Place, Hilo, HI 96720, USA}
\email[show]{aurora.mata.san@gmail.com}  

\author[0000-0002-6822-2254]{Emanuele Paolo Farina}
\affiliation{International Gemini Observatory/NSF NOIRLab, 670 N A`ohoku Place, Hilo, HI 96720, USA}
\affiliation{INAF -- Osservatorio di Astrofisica e Scienza dello Spazio di Bologna, via Gobetti 93/3, I 40129, Bologna, Italy}
\email{emanuele.paolo.farina@gmail.com}  

\author[0009-0009-8274-441X]{Anniek J.\ Gloudemans}
\affiliation{NSF NOIRLab, Gemini Observatory, 670 N A`ohoku Place, Hilo, HI 96720, USA}
\email{anniek.gloudemans@noirlab.edu}

\author[0000-0002-3007-0013]{Debora Pelliccia}
\affiliation{UCO/Lick Observatory, Department of Astronomy \& Astrophysics, UC Santa Cruz, 1156 High Street, Santa Cruz, CA 95064, USA}
\email{}

\author[0000-0002-2662-8803]{Roberto Decarli}
\affiliation{INAF -- Osservatorio di Astrofisica e Scienza dello Spazio di Bologna, via Gobetti 93/3, I 40129, Bologna, Italy}
\email{}

\author[0000-0002-2931-7824]{Eduardo Ba\~nados}
\affiliation{Max-Planck-Institut f\"ur Astronomie, K\"onigstuhl 17, D-69117, Heidelberg, Germany}
\email{}

\author[0000-0003-4747-4484]{Silvia Belladitta}
\affiliation{Max-Planck-Institut f\"ur Astronomie, K\"onigstuhl 17, D-69117, Heidelberg, Germany}
\affiliation{INAF -- Osservatorio di Astrofisica e Scienza dello Spazio di Bologna, via Gobetti 93/3, I 40129, Bologna, Italy}
\email{}

\author[0000-0002-4314-021X]{Manuela Bischetti}
\affiliation{IFPU, Institute for Fundamental Physics of the Universe, Via Beirut 2, 34014 Trieste, Italy}
\affiliation{INAF -- Osservatorio Astronomico di Trieste, Via G. B. Tiepolo 11, I-34131 Trieste, Italy}
\email{}

\author[0000-0001-8582-7012]{Sarah E.\ I.\ Bosman}
\affiliation{Max-Planck-Institut f\"ur Astronomie, K\"onigstuhl 17, D-69117, Heidelberg, Germany}
\email{}

\author[0000-0001-9488-238X]{Xander Byrne}
\affiliation{Institute of Astronomy, University of Cambridge, Madingley Road, Cambridge CB3 0HA, UK}
\email{}

\author[0000-0002-7898-7664]{Thomas Connor}
\affiliation{Center for Astrophysics $\vert$\ Harvard\ \&\ Smithsonian, 60 Garden St., Cambridge, MA 02138, USA}
\email{}

\author[0000-0003-0821-3644]{Frederick Davies}
\affiliation{Max-Planck-Institut f\"ur Astronomie, K\"onigstuhl 17, D-69117, Heidelberg, Germany}
\email{}

\author[0000-0001-6179-7701]{Thales A.\ Gutcke}
\affiliation{Institute for Astronomy, University of Hawai`i, 2680 Woodlawn Drive, Honolulu, HI 96822, USA}
\email{}

\author[0000-0003-3762-7344]{Weizhe Liu}
\affiliation{Steward Observatory, University of Arizona, 933 N. Cherry Ave., Tucson, AZ 85721, USA}
\email{}

\author[0000-0002-5941-5214]{Chiara Mazzucchelli}
\affiliation{Instituto de Estudios Astrof\'{i}sicos, Facultad de Ingenier\'{i}a y Ciencias, Universidad Diego Portales, Avenida Ej\'ercito Libertador 441, Santiago, Chile}
\email{}

\author[0000-0001-5492-4522]{Romain Meyer}
\affiliation{Department of Astronomy, University of Geneva, Chemin Pegasi 51, 1290 Versoix, Switzerland}
\email{}

\author[0009-0009-1715-4157]{Silvia Onorato}
\affiliation{International Gemini Observatory/NSF NOIRLab, 670 N A`ohoku Place, Hilo, HI 96720, USA}
\email{}

\author[0000-0003-4793-7880]{Fabian Walter}
\affiliation{Max-Planck-Institut f\"ur Astronomie, K\"onigstuhl 17, D-69117, Heidelberg, Germany}
\email{}

\author[0000-0002-7633-431X]{Feige Wang}
\affiliation{Department of Astronomy, University of Michigan, 1085 S. University Ave., Ann Arbor, MI 48109, USA}
\email{}

\author[0000-0001-5287-4242]{Jinyi Yang}
\affiliation{Department of Astronomy, University of Michigan, 1085 S. University Ave., Ann Arbor, MI 48109, USA}
\email{}

\begin{abstract}
We present a detailed multi-wavelength analysis of two gravitationally lensed $z>6$ quasar candidates: J0109--5424 and P170+20. Using JWST/NIRSpec near-infrared data from the \textit{Aether} survey, we demonstrate that J0109--5424 is in fact a FeLoBAL quasar at $z=2.07$, exhibiting strong iron broad absorption lines that mimic the Lyman break. From the \pab{} emission line, we estimate a black hole mass of $3.9\times10^8$\,M$_{\odot}$ and a bolometric luminosity of $L_{\text{bol}} = 7.7 \times 10^{45}\,\text{erg s}^{-1}$ for this quasar.
Follow-up HST/ACS WFC imaging of P170+20 with the F555W filter, probing emission below the Lyman limit of the quasar, reveals a compact source located $0.09\pm0.04\arcsec$ from the quasar position. PSF modelling and subtraction yield no evidence for extended emission. This is inconsistent with a foreground galaxy acting as a lens. Furthermore, we obtain \mbox{Gemini-South/Flamingos 2} near-infrared observations to complement the existing optical Gemini-North/GMOS spectrum of P170+20. Our spectral analysis shows that the observed spectral shape of P170+20 is consistent with a composite quasar and M-dwarf spectrum. However, this interpretation remains uncertain, as a chance alignment between an ultracool dwarf and a quasar within $0.1\arcsec$ would be unlikely. Overall, our findings establish J0109--5424 as an intermediate redshift source and do not favor the lensed nature of P170+20, although this possibility cannot be conclusively ruled out, highlighting the challenges of selecting lensed quasars at high redshift.
\end{abstract}

\keywords{\uat{quasars: individual (J0109--5424, P170+20)}{}  --- \uat{gravitational lensing: strong} {} --- \uat{quasars: supermassive black hole}{}}

\section{Introduction}\label{sec:introduction}

High-redshift luminous quasars ($z > 5$) serve as powerful probes of the first billion years of the Universe, providing crucial insight into early structure formation, the assembly of massive galaxies and supermassive black holes (SMBHs), and the physical conditions of the intergalactic medium \citep[e.g.,][]{Fan2023}. To this end, the identification of high-redshift quasars has increased dramatically over the past two decades, driven by advances in several multi-wavelength surveys, including the Sloan Digital Sky Survey \citep[SDSS,][]{York2000}, the VISTA Kilo-Degree Infrared Galaxy survey \citep[VIKING, e.g.,][]{Aranboldi2007}, the Canada-France High-z Quasar Survey \citep[CFHQS,][]{Omont2013}, the Panoramic Survey Telescope and Rapid Response System \citep[Pan-STARRS1,][]{2016arXiv161205560C}, the Hyper Suprime-Cam Subaru Strategic Program \citep[HSC-SSP,][]{Aihara2018}, the DESI Legacy Imaging Surveys \citep[DELS,][]{Dey2019AJ....157..168D}, and the Euclid mission \citep{Euclid2025}. These efforts have led to the discovery of hundreds of high-redshift quasars to date \citep[e.g.,][]{Fan2001, Willott2010discovery, Venemans2013, Matsuoka2016, Gloudemans2022, Yang2023, Banados2023, Banados2025, Belladitta2025}.

The discovery of high-redshift quasars is challenging due to their low spatial density and contamination from ultracool dwarfs with similarly red colors. Consequently, they are commonly selected using optical and near-infrared (NIR) color criteria \citep[e.g.][]{Fan2001, Yang2019b}. A key feature exploited in these selections is the pronounced spectral break between adjacent photometric bands \citep[e.g., the $r$- and $i$-bands for $z>5$][]{Fan2023}, caused by near-complete absorption at rest-frame wavelengths below 1215.67\,\AA{} by the neutral intergalactic medium during reionization.
Nevertheless, traditional color-selection techniques are inherently limited and may exclude certain quasar populations. In particular, strongly gravitationally lensed quasars can be easily missed, as the requirement of non-detection in blue bands essentially removes high-redshift quasars aligned with foreground sources \citep[e.g.,][]{McGreer2010, WangFeige2019LF}.

Identifying lensed quasars at high redshift is crucial.
Much of our current knowledge of high-redshift quasars relies on the assumption that their observed luminosities are intrinsic to the sources.
The magnification of quasar luminosities by gravitational lensing can lead to an overestimation of their black hole masses (M$_{\mathrm{BH}}$) if lensing is not properly accounted for \citep[e.g.,][]{Comerford2002, Fan2019} and can bias the bright end of the quasar luminosity function \citep[e.g.,][]{Venemans2013, WangFeige2019LF, Pacucci2022, Schindler2023}.
At the same time, this phenomenon offers valuable insights into the intrinsic properties of high-$z$ quasars. Lensing magnification provides a unique opportunity to probe quasar properties in greater detail, since the stretching of lensed images yields higher effective spatial resolution and improved separation between the central quasar and its host galaxy \citep[e.g.,][]{Peng2006, Yang2019a}.
In addition, gravitational lenses can place constraints on the Hubble constant through measurements of time delays between multiple quasar images \citep[e.g.,][]{Suyu2017, Millon2020}, and they can be used to investigate the properties of dark matter \citep[e.g.,][]{Gilman2020, Yue2023}. 

Dedicated searches for gravitational lenses have been extensive, and the use of automated algorithms and machine learning techniques has enabled the identification of numerous high-redshift candidates \citep[e.g.,][]{Aihara2022, Yue2023, Andika2023}.
Lensed quasars are typically selected using a combination of photometry, to determine the positions and flux densities of resolved components, and spectroscopy, to confirm the nature of both the quasar and the lensing galaxy. However, this approach can sometimes instead lead to the discovery of quasar pairs, as quasars separated by $\sim$kpc scales may exhibit similar angular separations ($\Delta \theta < 1^{\prime\prime}$) and colors \citep[e.g.,][]{Yue2023}.
Despite these efforts, the only confirmed gravitationally lensed quasar at $z>6$ is J0439+1634 at $z = 6.51$ \citep{Fan2019}. Its luminosity is amplified by gravitational lensing, with a magnification factor of $\mu \sim 50$, making it the brightest source currently known at $z>5$.

Recently, \citet{Byrne2024} applied an unsupervised machine learning algorithm to DES imaging to identify high-$z$ quasar candidates, enabling the selection of sources, including gravitationally lensed quasar candidates, previously excluded by traditional color criteria that require non-detections in the $g$ or $r$ bands. This approach led to the discovery of the $z = 6.07$ lensed quasar candidate J0109--5424, which exhibits strong broad absorption line (BAL) features. Gemini-South/GMOS spectroscopy revealed significant flux throughout the Ly$\alpha$ forest and emission below the Lyman limit (i.e., in a spectral range where the quasar emission should be virtually zero), interpreted as a foreground source and supporting a lensing scenario.
Another promising lensed quasar candidate, P170+20 at $z = 6.41$, was serendipitously discovered by \citet{Banados2023} in the Pan-STARRS1 survey using standard color-selection criteria. P170+20 was reported to exhibit a peculiar spectrum, characterized by the absence of strong emission lines, a sharp flux break at an observed wavelength of 8992\,\AA{}, and excess flux blueward of the Lyman break that declines smoothly. This behaviour is inconsistent with high transmission spikes in the \lya{} forest and with the emission detected in the DELS $r_{\mathrm{DE}}$-band [$r_{\mathrm{DE}} = (23.11 \pm 0.09)$\,mag]. The presence of this flux excess suggests a foreground source, making P170+20 another promising candidate for a gravitationally lensed high-$z$ quasar.

In this paper, we present space- and ground-based follow-up observations of both J0109--5424 and P170+20 to confirm or refute their gravitationally lensed nature. The paper is structured as follows. In \autoref{sec:observations}, we summarize the observations and data reduction procedures. In \autoref{sec:J109--5424} and \autoref{sec:P170+20}, we present the spectral analyses and provide evidence against the gravitationally lensed nature of both quasars. Finally, the main findings are summarized in \autoref{sec:summary}.
All magnitudes reported in this paper are given in the AB system \citep{Oke1974, Oke1983}. We adopt a $\Lambda$CDM cosmology with $H_0 = 70\,\kms\,\text{Mpc}^{-1}$, $\Omega_{\mathrm{M}} = 0.3$, and $\Omega_{\Lambda} = 0.7$. In this cosmology, at $z = 6$ the angular scale of 1\arcsec{} corresponds to 5.71\,kpc.

\section{Observations and Data Reduction} \label{sec:observations}

Both J0109--5424 and P170$+$20 have been followed up with multiple optical and NIR facilities, including:
the Hubble Space Telescope (HST) Advanced Camera for Surveys \citep[ACS,][]{Ryon2022};
the Gemini South and North Multi Object Spectrographs \citep[GMOS,][]{Hook2004, Gimeno2016, Scharwaechter2017};
the Gemini South Flamingos-2 imager and spectrograph \citep{Eikenberry2006};
The Large Binocular Telescope (LBT) Utility Camera in the Infrared \citep[LUCI,][]{Seifert2003, Ageorges2010};
and
the James Webb Space Telescope (JWST) Near Infrared Spectrograph \citep[NIRSpec,][]{Jakobsen2022, Boker2022}.
An overview of the observations used in this work is shown in \autoref{tab:observations}, which are described in detail in the following sections. 

\begin{table*}
     \begin{tabular}{llccc}
     \toprule
     Telescope - Instrument     & Obs.\ date & Filter/Grating & Wavelength range [nm]                  & Exp.\ time [s]\\
     \midrule
     \multicolumn{5}{l}{\textbf{J0109$-$5424} (RA=01:09:09.0; Dec=$-$54:24:16.29)}\\
     \quad Gemini South - GMOS        & 2022-12-11 & GG455/R400    & \phantom{1}520 -- 1035                 & 2000      \\
     \quad JWST - NIRSpec/IFU         & 2024-08-27 & F290LP/G395H  & 2870 -- 5270                           & 2600      \\
     \midrule
     \multicolumn{5}{l}{\textbf{P170$+$20\phantom{00}} (RA=11:23:19.83; Dec=$+$20:12:29.75)}                            \\  
     \quad HST - ACS/WFC              & 2023-07-12 & F555W/\nodata       & \phantom{1}458 -- \phantom{1}621 & 2080      \\
     \quad Gemini North - GMOS        & 2022-06-13 & RG610/R831          & \phantom{1}715 -- \phantom{1}973 & 2400      \\
     \quad LBT - LUCI1/LUCI2 & 2026-03-25 & zJspec/G200               &           900 -- 1200           & 3600 \\
     \quad Gemini South - Flamingos~2 & 2023-05-25 & HK/HK               &           1228 -- 2595           & 6000      \\
     \bottomrule
     \end{tabular}
     \caption{Overview of the observations collected for the two quasars, J0109$-$5424 and P170$+$20.}
     \label{tab:observations}
\end{table*}

\subsection{Observations of J0109--5424}

\paragraph{Gemini South/GMOS}
The quasar discovery spectrum of J0109--5424 was obtained with Gemini South/GMOS as part of program GS-2022B-FT-208 (PI: Farina). The GMOS detector was binned by 2 pixels in the spectral and 2 pixels in the spatial directions. The instrument was configured with the GG455 order-blocking filter, the R400 grating, and the 1.5\arcsec{} wide slit delivering a nominal spectral resolution $R=\lambda$/$\Delta \lambda\sim650$ in the 5200\,\AA{}-10350\,\AA{} wavelength range. We collected $4\times500$\,s exposures with small spatial and wavelength dithers between them.
Data has been reduced following standard procedures using \textsc{Pypeit}
\citep{Prochaska2019pypeit2, prochaska2020pypeit}.
Finally, the spectrum was rescaled to match the Pan-STARRS1 $z$-band  magnitude of the source $z_{\rm PS1} = 20.18$\,mag. 
See also \citet{Byrne2024} for further details on the GMOS discovery spectrum.  

\paragraph{JWST/NIRSpec}
Data have been collected with JWST/NIRSpec in the IFU configuration as part of the Cycle 3 JWST Survey \#5645 dubbed the \textit{Aether survey} \citep[][, Farina et al.\ in prep.]{Farina2024}.
The instrument was configured with the G395H/F290LP grating, delivering $R\sim2700$ between 2.87\,$\mu$m and 5.27$\mu$m.
Integrations were set to use the {\texttt{NRSINRS2}} readout pattern with 9~Groups, 1~Integration, and 4 {\texttt{SMALL CYCLING}} dither positions for a total time of $\sim$45 min on target.
The data reduction has been performed using the JWST pipeline developed by the Space Telescope Science Institute following the procedure described in \citet{Loiacono2024} and \citet{Decarli2024}.
The quasar spectrum was extracted from the NIRSpec IFU data using a circular aperture with a radius of 3 pixels (0.3\arcsec) centred on the quasar, with local background subtraction applied. The resulting spectrum was aperture-corrected using the correction factors derived by \cite{Loiacono2024}.

\subsection{Observations of P170+20}

\paragraph{HST/ACS/WFC}
Imaging of the field of P170$+$20 has been collected with the ACS/WFC instrument on HST as part of the GO program \#17284 \citep[PI: Farina,][]{Farina2022}. 
Data have been collected with the F555W filter performing 4$\times$520s exposures with the \texttt{ACS-WFC-DITHER-BOX} pattern, resulting in a total time on target of 35\,min.
Processed images has been downloaded from the Barbara A.\ Mikulski Archive for Space Telescopes. 

\paragraph{Gemini North/GMOS}
The high redshift nature of P170$+$20 was confirmed with the Gemini North/GMOS program GN-2022A-Q-411 (PI: Farina).
The target was observed for $4\times600$\,s exposures with spatial and spectral dithers one from the other. 
Data has been taken with the R831 grating, the RG610 order blocking filter, and the 1.5\arcsec{} wide slit, covering roughly the  3600\,\AA{}--9735\,\AA{} spectral range with nominal resolution $R\sim1465$.
The reduction has been performed with \textsc{Pypeit}, following standard procedures and an improved treatment of the flux calibration which takes into account the different response between the three GMOS' CCDs \citep[][]{Scharwaechter2017}.
Absolute flux calibration was obtained by rescaling the data to the $z$-band magnitude $z_{\rm PS1} = 20.38$\,mag. 
For additional details on the discovery spectrum see \citet{Banados2023}.

\paragraph{Gemini South/Flamingos~2}
NIR spectroscopy of P170$+$20 was collected with Gemini South/Flamingos~2 as part of the program GS-2023A-FT-209 (PI: Farina) using the 3-pixels (0.54\arcsec{}) wide slit, the HK grism and spectroscopic filter.
We collected 5$\times$ABBA sequences with 300\,s per exposure, resulting in a total time on target of 100\,min.
Data has been reduced with \textsc{Pypeit} and the resulting spectrum fully covers the $H$ and $K$-band with a resolution of $R\sim900$.
We also reduced the $H$-band field image used for acquisition in order to estimate the magnitude of the source in this band. 
The two 10\,s frames has been processed using standard recipes part of the \textsc{DRAGONS} pipeline \citep[][]{Labrie2019, Labrie2023} and the photometric zero-point was determined using the 2MASS point source catalogue \citep[][]{Cutri2003, Skrutskie2006}.
The Flamingos~2 spectrum has been rescaled to match the measured $H$-band magnitude of the source $H = (19.70 \pm 0.16)$\,mag.

\paragraph{LBT/LUCI1 and LUCI2} P170+20 was observed with LUCI at the LBT.
The observations were carried out in binocular mode using both LUCI1 and LUCI2 as part of the program MPIA-2026A-001 (PI: Ba{\~n}ados). We employed the G200 grism in combination with the zJspec filter, providing spectral coverage over the ($\sim0.9{-}1.2\,\mu\mathrm{m}$) wavelength range, which includes the expected position of the \civ{} emission line. Long-slit spectroscopy was performed using a slit width of 1.0\arcsec{}, and an ABBA dithering pattern was adopted to ensure optimal sky subtraction. The total on-source integration time was 1 hour, consisting of individual exposures of 240\,s each. The data was reduced using \textsc{Pypeit}. The reduction procedure included standard processing steps such as bias subtraction, flat-field correction, wavelength calibration using arc-lamp exposures, and flux calibration based on spectrophotometric standard stars.  The final 1-D spectrum was normalized to the PanSTARRS $z$-band magnitude. 

\section{J0109--5424: A FeLoBAL quasar at z=2.07}\label{sec:J109--5424}

\begin{figure*}
\centering
\includegraphics[width=1.00\textwidth]{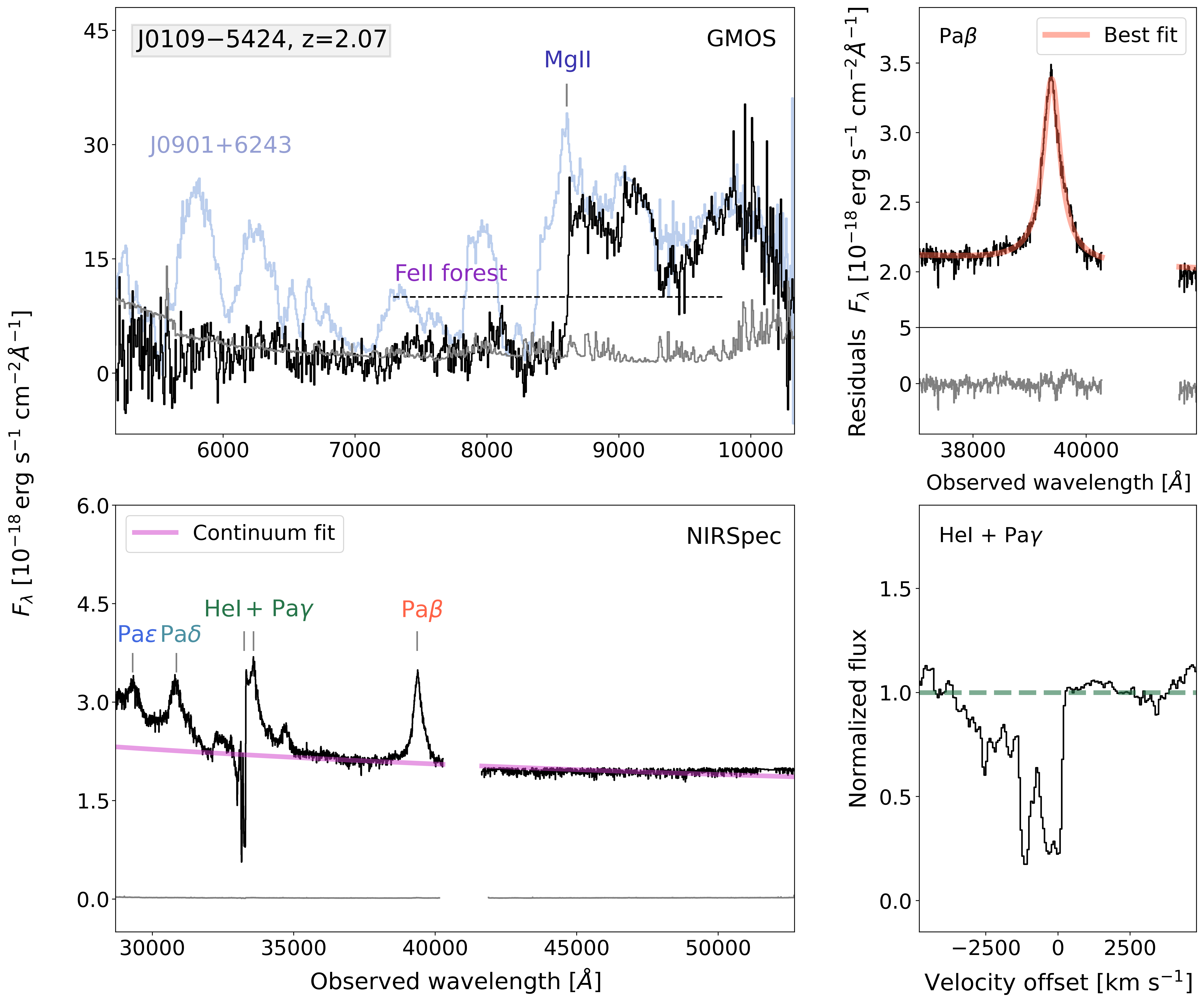}
\caption{
Quasar spectra of J0109--5424 obtained with GMOS and NIRSpec.
\textit{Upper left Panel}: GMOS optical spectrum (black line) and associated error (gray line), with the \mgii{} and iron absorption features disguised as a \lya{} break at 8604\,\AA{} labeled. The FeLoBAL J0901+6243 spectrum is over-plotted in light blue for comparison.
\textit{Lower left Panel}: NIRSpec IFU spectrum (black line) and associated error (gray line), with the Paschen-series emission lines redshifted to $z = 2.07$ marked. The best-fit to the continuum is shown as a magenta line.
\textit{Upper right Panel}: Zoom-in of the \pab{} emission-line region. The best-fit model is shown in light red, and the residuals are displayed in the lower panel in gray.
\textit{Lower right Panel}: Normalized flux of the \hei{} and \pag{} complex. The green dotted line indicates the normalized flux level of one as a reference. The velocity offset is computed relative to the redshift derived from the \pab{} line. The BAL feature extends to velocities $\lesssim -2500\, \kms$ relative to the quasar.
}
\label{fig:J0109spec}
\end{figure*}

The resulting GMOS (re-binned to 6.65\,\AA{} bins) and NIRSpec IFU spectra of J0109--5424 are shown in the upper-left and lower-left panels of \autoref{fig:J0109spec}, respectively. The NIRSpec IFU spectrum displays a series of strong emission lines that we identify as the Hydrogen Paschen series (specifically \pab{}, \pag{}, \pad{}, and \pae{}) at a redshift of $z = 2.07$. 
Our observations therefore demonstrate that the break detected by \citet{Byrne2024} at $\sim8604$\,\AA{} is not the \lya{} break at $z \sim 6$. 
Instead, we interpret the sharp break as a series of blended \mgii{} and iron broad absorption lines (BALs) in the quasar spectrum.
Indeed, BAL quasars often distort the profiles of prominent emission lines used to determine redshifts, as absorption depresses the blue wings of one or more spectral features \citep{Filbert2023}. 
Together with the absence of strong emission lines in the observed optical wavelength range, this supports the classification of J0109--5424 as a FeLoBAL quasar.
For comparison, the spectrum of the FeLoBAL quasar J0901+6243 was retrieved from SDSS, shifted to the redshift of J0109--5424, and rescaled to match the median flux of J0109--5424 over 9800--10000\,\AA{}. Compared with J0901+6243, J0109--5424 shows more extreme properties, remaining unusually flat blueward of the \mgii{} line, rather than becoming brighter again toward shorter wavelengths.
The spectral properties of such objects are commonly attributed to accretion-disk winds and/or a blowout phase in ultra-luminous infrared galaxies (ULIRGs) that expels gas and dust \citep[e.g.,][]{Choi2022III, Choi2022}. FeLoBAL quasars are often missed in optical surveys because of their reddened spectra and lack of strong emission lines; however, they can contaminate $z > 5.7$ quasar searches, as they occupy a similar locus in optical color space. 

\subsection{Black Hole Mass \& Accretion Rate}

We performed spectral modelling and analysis using the software tool \textsc{Sculptor}\footnote{\url{https://github.com/jtschindler/sculptor}} \citep{Schindler2022}, which was specifically developed to fit the emission of high-$z$ quasars  \citep[e.g.,][]{Schindler2020, Schindler2021, Farina2022MBH}. The estimates of the black hole mass and accretion rate of J0109--5424 were derived from fitting the high S/N \pab{} emission line.
We first modeled the quasar continuum with a simple power law, using selected wavelength regions free of significant emission-line contamination (36000–38400\,\AA, 42000–44000\,\AA, and 47500–50000\,\AA). The resulting continuum exhibits a relatively flat power-law slope of $-0.5$ (magenta line in \autoref{fig:J0109spec}) and was subtracted from the observed spectrum. The \pab{} line was then fitted with a single Lorentzian component centred at $\lambda = 39404$\,\AA. The fitting result is shown in light red in the upper-right panel of \autoref{fig:J0109spec}.

From the fit to the \pab{} line, we precisely derive the redshift of J0109--5424 to be $z = 2.0741 \pm 0.0002$.
In addition, we estimate the black hole mass using the FWHM ($\mathrm{FWHM}_{Pa\beta}$) and the luminosity ($L_{Pa\beta}$) of the \pab{} line \citep[see][]{Kim2015}: 
\begin{equation}
    \frac{M_{\text{BH},Pa\beta}}{M_{\odot}}=10^{7.04} \left(\frac{L_{Pa\beta}}
    {10^{42}\, \text{erg~s}^{-1}}\right)^{0.48}
    \left(\frac{\text{FWHM}_{Pa\beta}}{10^3 \,\text{km s}^{-1}}\right)^2,
\end{equation}
which has an intrinsic scatter of $\sim$0.2\,dex.
The derived luminosity $L_{\mathrm{Pa}\beta} = 2.35 \times 10^{43}\,\text{erg s}^{-1}$ and $\mathrm{FWHM}_{Pa\beta} = 2792\,\text{km s}^{-1}$ yield a black hole mass of $M_{\text{BH}} = 3.9 \times 10^8\,M_{\odot}$.
The broad, Lorentzian profile of the \pab{} line indicates its association with the quasar broad line region rather than a potential foreground lens.
Following \citet{Kim2015}, we also estimate the bolometric luminosity of J0109--5424 from the \pab{} line luminosity:
\begin{equation}
    \log \left(\frac{L_{\text{bol}}}{10^{44}\text{erg~s}^{-1}}\right)= 1.29 + 0.969~\log\left(\frac{L_{Pa\beta}}{10^{42}\text{erg~s}^{-1}}\right),
\end{equation}
which yields $L_{\text{bol}} = 7.7 \times 10^{45}\,\text{erg s}^{-1}$.
Finally, the Eddington ratio ($\lambda_{\text{Edd}}$), defined as the ratio between the bolometric luminosity of the quasar and its Eddington luminosity \citep{Eddington1926}, is $\lambda_{\text{Edd}} = 0.16$.
J0109--5424 is thus one of the highest-redshift FeLoBAL quasars known to date \citep{Ross2015} with its black hole properties being consistent with those of the general $z < 1$ FeLoBAL quasar population reported by \citet{Leighly2022}.
Although our observations show no clear evidences for gravitational lensing, the revised redshift alone does not entirely exclude that the $z\sim2$ quasar might still be lensed. However, a potential foreground lensing galaxy would necessarily lie at $z<2.07$ and might be detectable in optical imaging.

\subsection{BAL velocity offset}

The NIRSpec IFU spectrum of J0109--5424 shows a clear absorption feature blueward of the \pag{} line, which we identify as \hei{} at 1.0830\,$\mu$m in absorption. Because this feature is not blended with multiple components or species, as is the case for the \mgii{} plus iron complex, it can be used to constrain the velocity components that constitute the broad absorption line gas in this quasar.

To investigate this absorption, we model the \pag{} and \hei{} complex using two Gaussian components.
As the \hei{} line is almost completely blanked by the absorbing medium, we constrained it to share the same redshift and FWHM of the \pag{} line. The lower-right panel of \autoref{fig:J0109spec} shows the \pag{} + \hei{} complex divided by the best-fitting continuum and line model. The velocity offset is computed relative to the redshift derived from \pab{}.
The blueshift of the \hei{} absorption line indicates the presence of a strong quasar outflow. Although the intrinsic limitations of our modelling of the \pag{} and \hei{} lines preclude a more detailed analysis, we find that the absorbing gas extends to velocities of at least $-2500$\,km\,s$^{-1}$. The absorption appears to consist of at least two major components: one located just blue-ward of the systemic redshift and another offset by $\sim$1200\,km\,s$^{-1}$, and additional features extending out to $\sim 2500\,\kms{}$, similar to what has been observed in other FeLoBAL quasars \citep[e.g.,][]{Choi2022III}.

\section{P170+20: a potential quasar-star chance alignment} \label{sec:P170+20}

While the $z \sim 2$ nature of J0109--5424 is evident, our current data do not allow us to draw a similarly definitive conclusion regarding the lensing nature of P170$+$20. In the following section, we present several lines of evidence supporting the hypothesis that this system is not a lensed quasar but instead consists of a close alignment of a high redshift quasar and a compact foreground source, potentially an M-dwarf star.
We also discuss several caveats that leave the gravitational lensing scenario still as a viable possibility.

\subsection{Evidence against the lensing scenario}\label{sec:nolens}

\begin{figure*}
\centering
\includegraphics[trim={4.2cm 1.3cm 4.2cm 1.3cm}, clip, width=1.0\textwidth]{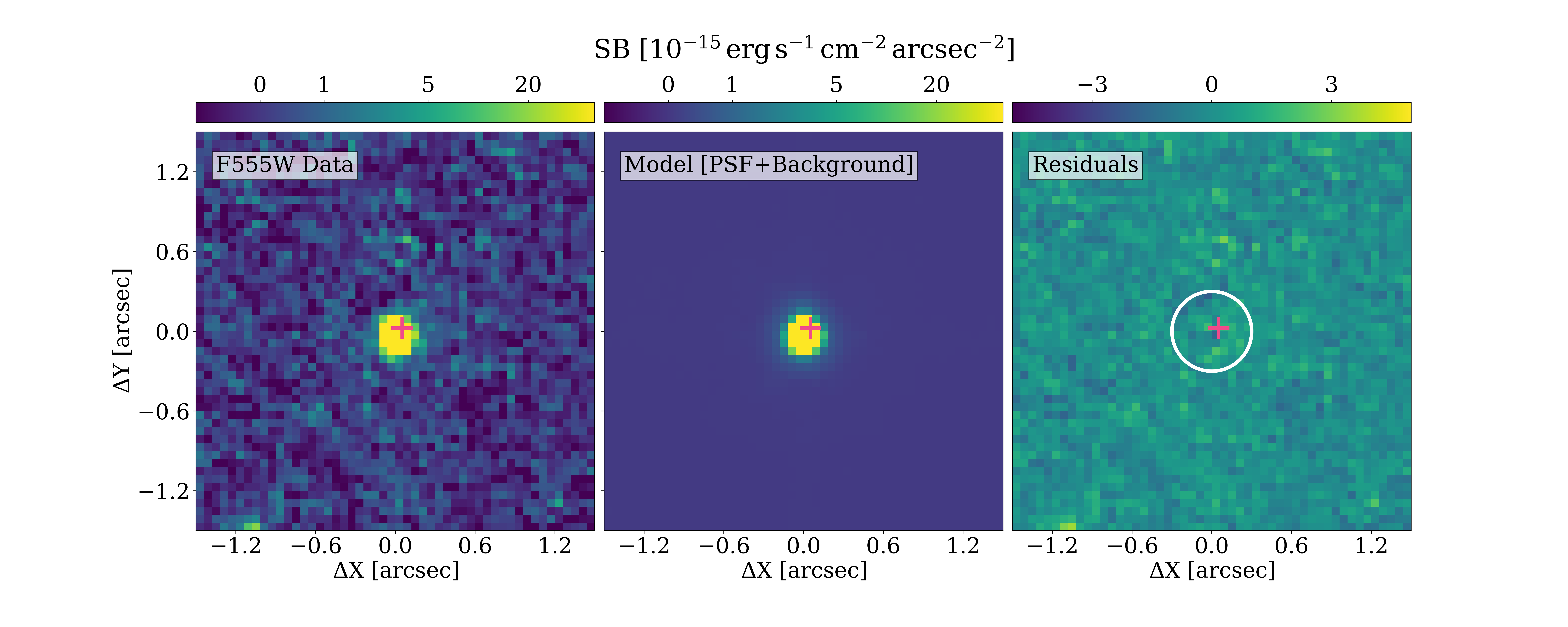}
\caption{GALFIT modelling of the foreground source of P170$+$20.
\textit{Left Panel}: HST ACS/WFC F555W image. The magnitude of the point source is $m_{F555W} = 24.34$\,mag.
\textit{Middle Panel}: Best-fit model image, consisting of a single PSF component with background subtraction.
\textit{Right Panel}: Corresponding residual map. The white circle marks an aperture with diameter of 0.5\arcsec{} centered on the source. The pink cross marks the DESI quasar position in all panels. Virtually no emission from the quasar is expected at these wavelengths as it is below the Lyman limit of the source.
}
\label{fig:Galfit}
\end{figure*}

\paragraph{Compactness of the foreground source}
Although massive galaxies at high redshift can have effective radii smaller than one kiloparsec \citep[e.g.,][]{vanDokkum2009}, the exquisite spatial resolution delivered by HST/ACS is generally sufficient to resolve galaxies at $z \sim 2 - 6$ \citep[e.g.,][]{Bouwens2004}. If P170$+$20 is indeed a high-$z$ lensed quasar, we would therefore expect the HST imaging to resolve the foreground lensing galaxy \citep[see, e.g.,][]{Fan2019}. The F555W filter used for the HST/ACS imaging covers the spectral range below the Lyman limit of the quasar, i.e., 4580 - 6210 \AA. Any detected emission in this band is thus expected to originate from the foreground source and should not be contaminated by the quasar itself.
The analysis of the HST ACS/WFC F555W image of P170$+$20 is shown in \autoref{fig:Galfit}. To test the point-source nature of the detected emission, we modeled the source with as a single PSF. Owing to the scarcity of bright stars in the field of P170$+$20, we were unable to construct the PSF directly from the science image. Instead, we used observations of the neutron star 1E 1207.4$-$5209 obtained with the same filter configuration as our dataset, on July 28th 2003 (programme \#9872).
This approach may limit our ability to achieve a perfect PSF fit due to the so-called HST ``breathing'', i.e., small focus variations caused, for example, by temperature changes in the telescope. However, such effects are expected to impact our analysis only at the level of a few percent \citep[][]{Decarli2012ApJ...756..150D, Bellini2018}.
We then used \textsc{GALFIT} \citep{Peng2002} to model the foreground-source emission with a PSF plus a flat background component. The best-fit PSF model and the corresponding residuals are shown in \autoref{fig:Galfit}. We find a compact source closely aligned with the quasar position.
The residuals within a circular aperture of $0.25\arcsec$ radius at the source position corresponds to a surface brightness of $\sim$27.9\,mag\,arcsec$^{-2}$, consistent with the background level of the image. We therefore find no evidence for extended emission in the residual image as it would be expected for a foreground lensing galaxy. Confirming that the source is instead well described as an unresolved point source.

\paragraph{Offset between background and foreground sources}
To first order, the angular separation between the foreground and background sources in a lensed system provides an upper limit
of the Einstein radius of the lens \citep[$\theta_{\mathrm{E}}$, see, e.g.,][]{Narayan1990MNRAS.243..192N, Chieregato2007A&A...474..777C}. If we assume that the lensing halo is well described by a singular isothermal sphere (SIS), $\theta_{\mathrm{E}}$ can be directly related to the velocity dispersion of the SIS ($\sigma_\mathrm{SIS}$), the angular diameter distance between the lens and the source ($D_{\mathrm{LS}}$), and the angular diameter distance between the observer and the source ($D_{\mathrm{S}}$):
$\theta_\mathrm{E}=4\pi\left(\sigma_\mathrm{SIS}/c\right)^2 D_\mathrm{SL}/ D_\mathrm{S}$.
We determined the offset between the quasar and the point source in the HST image using the quasar detection in the $z$-band image from the DESI Legacy Imaging Survey DR8. For both the Legacy Survey and the HST image, the astrometry is calibrated to Gaia Data Release~1 \citep{Gaia2016A&A...595A...2G}, with a systematic uncertainty of $0.03\arcsec$. The measured offset between the quasar and the point source is $0.09\arcsec \pm 0.04\arcsec$.
Taking this as an upper limit on $\theta_{\mathrm{E}}$  and adopting the Faber–Jackson relation for early-type galaxies presented in \citet{Nigoche2010} -- under the assumption that $\sigma_{\mathrm{SIS}} = \sigma_{\star}$ (i.e., that the SIS velocity dispersion equals the stellar velocity dispersion) -- and applying a $k$-correction derived from the elliptical and S0 galaxy templates of \citet{Mannucci2001}, we can directly relate the apparent magnitude of the source to its redshift. We can thus estimate that a foreground source with $m_{\mathrm{F555W}} = 24.34$\,mag would need to be located at a redshift of $z \lesssim 0.4$ to reproduce the observed configuration as a lens. A lower-redshift foreground source with a smaller Einstein radius could also produce some lensing while remaining consistent with the observed separation. At these relatively low redshifts, however, this would again imply that the lens galaxy should be resolved in the HST imaging \citep[e.g.,][]{vanDokkum2010ApJ...709.1018V}.

\begin{figure*}
\plotone{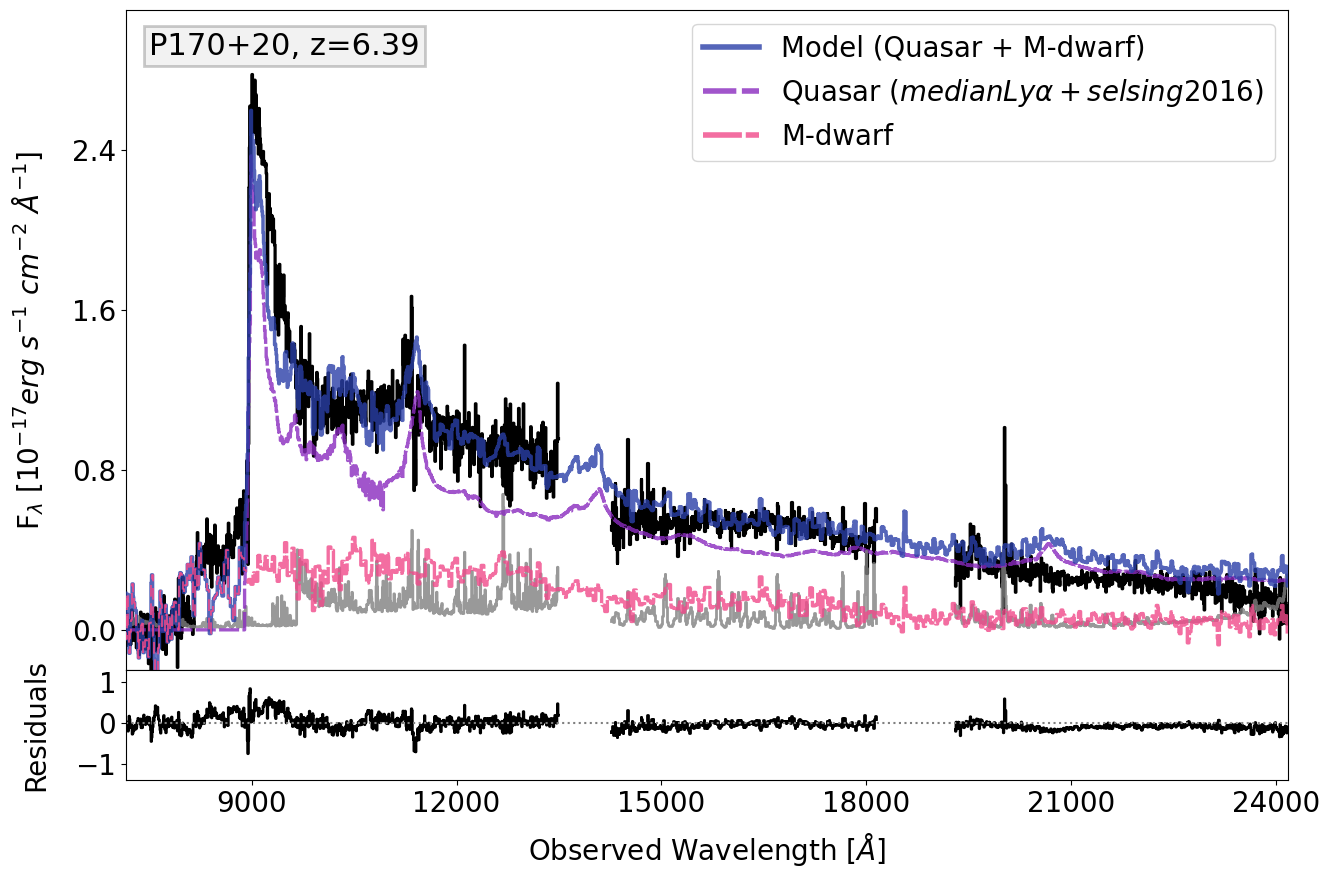}
\caption{Quasar spectrum of P170$+$20 obtained by combining the GMOS optical data, the LUCI1/LUCI2, and the Flamingos-2 near-infrared data.
\textit{Upper panel}: The observed spectrum is shown in black, with the associated error vector in gray. The best-fit model, as described in \autoref{sec:nolens}, is shown in blue. It consists of an M-dwarf spectrum displayed in pink and the \textit{median-\lya{}} + \textit{Selsing2016} quasar template shown in purple.
\textit{Lower panel}: Residuals of the best-fit model.}
\label{fig:P170+20spec}
\end{figure*}

\paragraph{Spectral analysis}
We analyze the spectral data of P170$+$20, consisting of Gemini GMOS optical and Flamingos-2 near-infrared, and LBT LUCI1/LUCI2 observations. The datasets are combined using a common wavelength grid with 7.7\,\AA\ bins. The resulting combined spectrum is shown in \autoref{fig:P170+20spec}.
We exclude the possibility that this source is a FeLoBAL quasar at z$\sim2$, given the absence of apparent \hb{} and \ha{} emission lines, which would be expected at $\sim1.6$ $\mu$m and $2.1$ $\mu$m, respectively, in the NIR spectra.
To avoid duplicate coverage in the wavelength region overlapping with GMOS, we excluded the blue end of the combined LUCI1/LUCI2 spectrum from the final spectrum.
Owing to the presence of strong telluric absorption bands in the Flamingos-2 data, we restrict our analysis in the near infrared to the wavelength ranges 14250–18150\,\AA\ and 19300–24200\,\AA. 
Since brown dwarfs are among the most common objects in the sky and represent a major source of contamination in quasar samples due to their faintness in optical bands, we will consider the possibility that the observed spectrum is a combination between an high-redshift quasar and a foreground brown dwarf as this scenario could explain the significant flux observed blue-ward of the \lya{} break at 8924\,\AA\ in the GMOS spectrum and the compactness of the source detected in HST imaging.
We constructed our model by using a combination of brown dwarf spectra and a quasar template. For the brown dwarf component, we selected spectra of M-, L-, and T-type brown dwarfs from the catalog provided by \citet{Burgasser2006}, reflecting progressively lower temperatures and changes in chemical composition \citep{Martin2017}.
For the quasar component, we used the three spectral templates provided by \citet{Banados2016} to describe the emission around the Ly$\alpha$ region of quasars at $z \sim 6$. These include the \textit{weak-\lya{}} template, defined as the median spectrum of the 10\% of quasars with the smallest rest-frame equivalent width of the \lya{} + \nv{} complex; the \textit{median-\lya{}} template, representing the median spectrum of all quasars in the sample; and the \textit{strong-\lya{}} template, defined as the median spectrum of the 10\% of quasars with the largest rest-frame equivalent width of the \lya{} + \nv{} complex.
Each of these templates was combined with the \textit{Selsing2016} template, derived from the median spectrum of seven bright quasars in the redshift range $1 < z < 2$ \citep{Selsing2016}, in order to extend the spectral coverage to longer rest-frame wavelengths. The templates were connected at 1479.5\,\AA, corresponding to the longest rest-frame wavelength covered by the \citet{Banados2016} templates, and were set to zero at wavelengths shorter than 8924\,\AA, which corresponds to the observed wavelength of the \lya{} emission line at $z = 6.39$\footnote{\citet{Banados2023} report a slightly higher redshift for this source ($z=6.41$) based on the comparison with quasar templates. The redshift of $z = 6.39$ better matches the break of the \lya{} line and it is consistent within uncertainties.}.
To build our models, we first scaled each brown dwarf spectrum to match the excess emission blue-ward of the \lya{} line (i.e., at wavelengths $\lambda < 8924$\,\AA, where the contribution from the quasar emission is expected to be minimal). We then subtracted the scaled brown dwarf spectrum from the original GMOS and Flamingos-2 data and used the residual to fit the combined quasar template, allowing both the normalization and redshift to vary. 
The derived best-fit model consists of a combination of an M-dwarf spectrum (2MASS J18112466$+$3748513) and the \textit{median-Ly$\alpha$} + \textit{Selsing2016} quasar template redshifted to $z = 6.39$. This result is shown in \autoref{fig:P170+20spec}, where the upper panel displays the combined GMOS and Flamingos-2 data in black and the best-fit composite spectrum constructed for P170$+$20 in blue. The contribution of the M-type brown dwarf spectrum is shown in pink, while the quasar model is shown in purple. Residual are shown in the bottom panel. 
We emphasize that our simple approach yields an overall reasonable model of the observed emission and involves only five free parameters: the relative intensity of the \lya{} + \nv{} emission of the quasar; the magnitude and redshift of the quasar; the spectral type of the brown dwarf; and the magnitude of the brown dwarf. Although a more sophisticated modeling approach would provide a more accurate description of the spectral emission, our results in \autoref{fig:P170+20spec} suggest that the quasar light is potentially contaminated by a stellar object located along the same line of sight. 

The alternative possibility that the foreground source is a compact lower-redshift galaxy is explored in \autoref{appendix}. We find that an elliptical-type galaxy at $z\approx1.04$ can reproduce the observed break at $\sim8900$\,\AA. However, the combined model does not provide an adequate fit to the near-infrared spectral shape and yields a relatively high $\chi^2$ value compared to the model employing a brown dwarf. This analysis therefore does not support the presence of a foreground galaxy.

\subsection{Evidence in favor of lensing }

\paragraph{``Too small'' a proximity zone}
The so-called \textit{proximity zones} are regions of enhanced transmission in the \lya{} forest near a quasar, produced by the intense ionizing radiation of the quasar that creates a locally ionized bubble in the surrounding intergalactic medium \citep{Cen2000, Haiman2001, Bolton2007}.
Generally, if the gas in the intergalactic medium is close to fully neutral -- as expected for $z>6$ quasars \citep[e.g.,][]{Bosman2022} -- the size of these proximity zones depends on the neutral fraction of the surrounding intergalactic medium, the emission rate of ionizing photons from the quasar (which is ultimately proportional to its UV luminosity), and the lifetime of the quasar.
Traditionally, the size of the proximity zone ($R_\mathrm{p}$) is defined as the distance in the \lya{} forest at which the quasar emission, smoothed over a scale of 20\,\AA, drops below 10\% of the continuum level \citep{Fan2006}. Given the contamination from the foreground object, we are unable to reliably derive this value for P170$+$20. Nonetheless, the GMOS spectrum shows a very sharp cutoff at the apparent break from the \lya{} forest -- potentially corresponding to a size as small as $R_\mathrm{p}\lesssim1$\,Mpc -- and no excess emission is detected in the corresponding \lyb{}.
Given the reported magnitude of the target \citep[$M_{1450}\sim-26.10$\,mag;][]{Banados2023}, the putative size of the proximity zone of P170$+$20 is a factor of $\sim3\times$ smaller than expected for similar quasars at these redshifts \citep{Eilers2017, Satyavolu2023, Onorato2025}. This may imply that P170$+$20 is extremely \textit{young}, with the onset of the current bright quasar phase occurring as recently as $10^3$–$10^4$ years ago \citep{Eilers2021}. Alternatively, given that $R_\mathrm{p}\propto10^{-0.4\mathrm{M}_{1450}/2.87}$ \citep[e.g.,][]{Onorato2025}, the observed size would be consistent with an intrinsically fainter quasar that has been magnified by a factor $\mu\gtrsim15$ \citep[see, e.g.,][]{Davies2020}. Thus supporting the lens scenario for this object.

\paragraph{Small probability of chance alignment}
The HST imaging and spectral analysis suggest the possible presence of an M dwarf at a projected separation of $0.09\arcsec$ from the quasar. To assess the likelihood of finding an ultracool dwarf at such a small separation, we estimate the expected number density of M-, L-, and T-dwarfs at the sky position of the quasar.
We follow the approach of \cite{Wagenveld2022}, which employs a Galactic model from \cite{Chen2001ApJ...553..184C} to derive the number densities of ultracool dwarfs in the Milky Way. The expected local number density is then obtained by integrating over an area of 1\,deg$^2$ and considering magnitude bins down to 25\,mag in the $z$-band. Given the spatial variation in number density across the sky, we evaluate the expected surface density at the location of 480 known high-$z$ quasars at $z \gtrsim 5.5$ from \cite{Fan2023}. The median of these surface densities yields an estimated ultracool dwarf density of $\sim 370$ deg$^{-2}$. For the projected offset between the quasar and the ultracool dwarf, we assume a typical ground-based $z$-band seeing of $1.0\arcsec$. Under these assumptions, the probability of a random chance alignment between a quasar and a ultracool dwarf is $\sim0.01\%$. This likelihood is sufficiently small to make a star–quasar chance alignment unlikely, thereby -- at least in principle -- supporting the lensing hypothesis. However, given that more than 480 quasars have now been identified at $z\gtrsim5.5$, the probability that at least one such system is aligned with a ultracool dwarf increases to $\sim4\%$. This test therefore does not fully rule out the possibility that the foreground source is a local star.

\section{Summary} \label{sec:summary}

In this work, we present a comprehensive multi-wavelength follow-up analysis of two recently proposed high-$z$ lensed quasar candidates, J0109$-$5424 and P170$+$20. With currently only one confirmed lensed quasar at $z>6$ \citep[J0439$+$1634;][]{Fan2019}, additional identifications are essential to improve our understanding of their abundance and astrophysical impact. J0109$-$5424 was previously identified by \citet{Byrne2024} as a high-redshift lensed quasar candidate, and P170$+$20 by \citet{Banados2023} on the basis of excess emission blueward of Ly$\alpha$. However, in this work we robustly rule out the interpretation of J0109$-$5424 as a lensed high-redshift quasar, while for P170$+$20 we find that lensing is unlikely, though it cannot be entirely excluded.

Our main findings regarding J0109$-$5424 are as follows:

\begin{enumerate}

    \item Using JWST/NIRSpec IFU data, we identify J0109$-$5424 as a FeLoBAL quasar, with a sharp break observed at $\sim$$8600$\,\AA\ caused by blended iron absorption features blueward of \mgii{}. This feature had previously been misidentified as a Lyman break based solely on Gemini South/GMOS data. 

    \item By fitting the high signal-to-noise \pab{} emission line, we obtain a refined redshift measurement of $z = 2.0747 \pm 0.0002$, making it one of the currently highest-redshift known FeLoBAL quasars.
    
    \item The fit to the \pab{} emission line further allows us to estimate the black hole mass, $M_{\text{BH}} = 3.9\times10^8\,\text{M}{_\odot}$, and its Eddington ratio, $\lambda_{\text{Edd}} = 0.16$, consistent with the properties of FeLoBAL quasars at lower redshifts.
    
\end{enumerate}

Our main findings regarding P170$+$20 are as follows:

\begin{enumerate}

    \item We obtained HST/WFC3 F555W imaging, covering the 458–621\,nm wavelength range below the Lyman break at $z\sim 6.4$, to search for a possible foreground source. We detect a point-like source with an offset of $\lesssim0.1\arcsec$ from the high-$z$ quasar. PSF modeling and subtraction confirm its point-like nature, with residuals consistent with the background noise. We therefore find no evidence for the presence of an extended foreground galaxy that could act as a lens. 

    \item We examined the previously published Gemini-North/GMOS spectrum together with a newly obtained Gemini-South/Flamingos-2 near-infrared spectrum. No clear emission lines are detected in the near-infrared spectrum. To account for the emission observed below the Lyman break, we consider two possible scenarios: \textit{(i)} an ultracool star aligned with the quasar position or, alternatively, \textit{(ii)} a compact foreground galaxy.  

    \textit{(i)} For the ultracool star scenario, the spectral shape is best fit by a quasar template combined with an M-dwarf template. However, this does not fully explain the observed break at $\sim8900$ \AA. An estimate of the number density of ultracool dwarfs and known high-$z$ quasars suggests the chance of finding such an alignment is on the order of $\sim$4\%, which makes it plausible. 

    \textit{(ii)} Alternatively, we model a quasar template combined with several galaxy templates spanning different redshifts. The spectral break at $\sim8900$\,\AA\, can be explained by an elliptical galaxy at $z\approx1.04$. However, this model does not properly explain the near-infrared spectral shape, and our HST imaging suggests this galaxy would be highly compact. We therefore favor the M-dwarf model. 

    \item The sharp Ly$\alpha$ break in the GMOS spectrum suggests P170$+$20 is either an extremely young quasar that did not have enough time to carve its proximity zone, or in fact a lensed quasar with a ``normal" size proximity zone. 

\end{enumerate}

Therefore, we conclude that J0109--5424 is unambiguously a FeLoBAL quasar at $z=2.07$. However, while a chance alignment between a $\boldsymbol{z=6.39}$ quasar and an M-dwarf star provides a plausible explanation for the observed properties of the source, further investigation is required to fully account for the emission detected below the \lya{} break in the spectrum of P170$+$20. A dedicated imaging analysis based on HST data, with a specific focus on the quasar emission, potentially split in different lensed components, could provide additional insight into the origin of the observed excess emission.
Alternatively, we can make the reasonable assumption that the brown dwarf is M6 type with an approximate J-band magnitude of 22\,mag, which is $\sim$2 magnitudes fainter than the quasar.
Under these conditions, the distance relation from \cite{Filippazzo2015ApJ...810..158F} suggests that it is located at a distance of $\sim$2.5 kpc. Given its latitude, the star would likely be in the Milky Way halo, where stars have typical velocities of $\sim$250 km s$^{-1}$ (e.g. \citealt{Koppelman2021A&A...649A.136K}). Therefore, imaging of an additional epoch could allow us to determine its proper motion. Our simple calculation suggests we need to wait $\sim$9 years to measure a 0.2\arcsec\ offset.

\begin{acknowledgments}

This work was enabled by observations made from the Gemini North telescope, located within the Maunakea Science Reserve and adjacent to the summit of Maunakea. We are grateful for the privilege of observing the Universe from a place that is unique in both its astronomical quality and its cultural significance.

A.M.S and E.P.F. are supported by the international Gemini Observatory, a program of NSF NOIRLab, which is managed by the Association of Universities for Research in Astronomy (AURA) under a cooperative agreement with the U.S. National Science Foundation, on behalf of the Gemini partnership of Argentina, Brazil, Canada, Chile, the Republic of Korea, and the United States of America.
This research is based on observations made with the NASA/ESA Hubble Space Telescope obtained from the Space Telescope Science Institute, which is operated by the Association of Universities for Research in Astronomy, Inc., under NASA contract NAS 5–26555. These observations are associated with the Cycle 30 HST program \#17284.
This work is based on observations made with the NASA/ESA/CSA James Webb Space Telescope. The data were obtained from the Mikulski Archive for Space Telescopes at the Space Telescope Science Institute, which is operated by the Association of Universities for Research in Astronomy, Inc., under NASA contract NAS 5-03127 for JWST. These observations are associated with the Cycle 3 JWST Survey \#5645.
Support for the HST program HST-GO-17284 and the JWST program JWST-GO-05645 was provided by NASA through grants from the Space Telescope Science Institute, which is operated by the Association of Universities for Research in Astronomy, Inc., under NASA contract NAS 5-03127.
This paper includes data from the LBT (Program ID: MPIA-2026A-001). The LBT is an international collaboration among institutions in the United States and Europe. At the time data were acquired for this research, LBT Corporation Members were the University of Arizona on behalf of the Arizona Board of Regents; Istituto Nazionale di Astrofisica, Italy; and The Ohio State University, representing OSU, University of Notre Dame, University of Minnesota, and University of Virginia.  This research used the facilities of the Italian Center for Astronomical Archives (IA2) operated by INAF at the Astronomical Observatory of Trieste.  Observations have benefited from the use of ALTA Center (alta.arcetri.inaf.it) forecasts performed with the Astro-Meso-Nh model. Initialization data of the ALTA automatic forecast system come from the General Circulation Model (HRES) of the European Centre for Medium Range Weather Forecasts.
C.M. acknowledges support from Fondecyt Iniciacion grant 11240336 and the ANID BASAL project FB210003.
R.D. acknowledges support from the INAF RF 2024 mini-grant  ``The interstellar medium at high redshift'' and from the PRIN MUR ``2022935STW'', RFF M4.C2.1.1, CUP J53D23001570006 and C53D23000950006.
T.C. acknowledges support from NASA Contract NAS8-03060 to the Chandra X-ray Center.
R.A.M. acknowledges support from the Swiss National Science Foundation (SNSF) through project grant 200020\_207349.
F.W. acknowledges support from NSF award AST-2513040.
We thank the anonymous referee for their valuable comments
and insights, which helped improve the quality of this
manuscript.
\end{acknowledgments}
Some of the data presented in this article were obtained from the Mikulski Archive for Space Telescopes (MAST) at the Space Telescope Science Institute. The specific observations analyzed can be accessed via \dataset[doi:10.17909/9awb-3772]{https://doi.org/10.17909/9awb-3772}

\facilities{HST(ACS), Gemini South(GMOS and Flamingos~2), Gemini North(GMOS), JWST(NIRSpec)}

\software{astropy \citep{2013A&A...558A..33A,2018AJ....156..123A,2022ApJ...935..167A},
          \textsc{DRAGONS} \citep{Labrie2019, Labrie2023}
          \textsc{Pypeit} \citep{Prochaska2019pypeit2,prochaska2020pypeit},
          \textsc{Sculptor} \citep{Schindler2022}
          }

\appendix

\section{Spectral analysis of P170+20 with a foreground galaxy model}
\label{appendix}

\begin{figure*}
\plotone{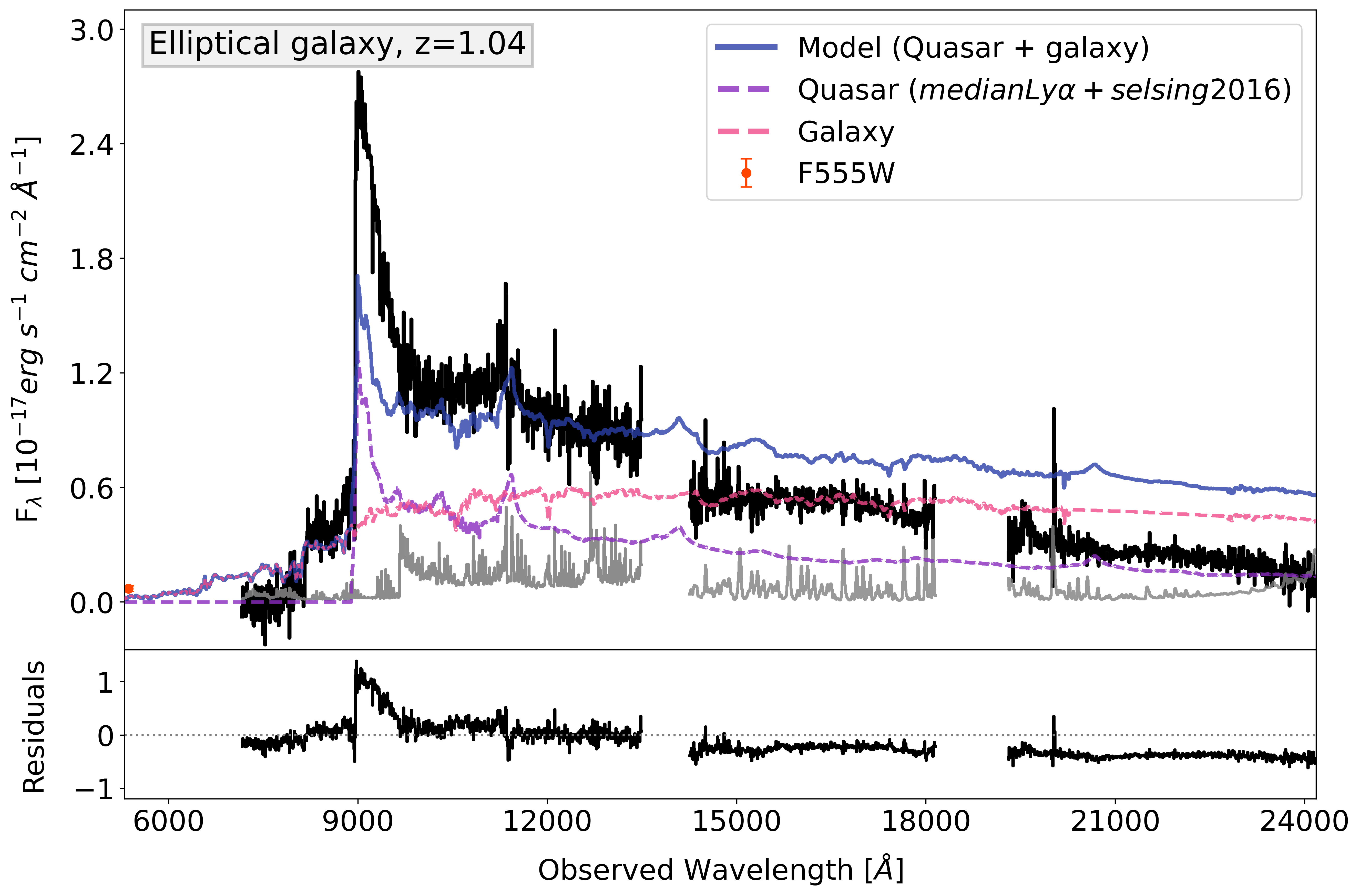}
\caption{Quasar spectrum of P170$+$20 obtained by combining the GMOS optical data, the LUCI1/LUCI2,  and the Flamingos-2 near-infrared data. \textit{Upper panel}: The observed quasar spectrum is shown in black, with the corresponding error vector in gray. The elliptical galaxy template is displayed in pink, the quasar template (\textit{median-\lya{}} + \textit{Selsing2016}) in purple, and the full combined model (quasar template + galaxy) in blue. The orange point indicates the F555W-band flux as a reference.  \textit{Lower panel}: Residuals of the fit.}
\label{fig:P170+20spec_galaxy}
\end{figure*}

As an alternative to the M-dwarf–quasar alignment proposed in \autoref{sec:nolens} for P170$+$20, we consider whether the compact source detected in the HST data could instead be a compact foreground galaxy. To test this hypothesis, we repeated our analysis of the combined GMOS–Flamingos-2 spectra using the Sa, Sb, Sc, S0, and E galaxy templates from \citet{Mannucci2001}, as well as the starburst galaxy templates sb1, sb3, and sb6 from \citet{Calzetti1994}.
The analysis followed the same procedure described in \autoref{sec:nolens}, with the additional step that the galaxy redshift was allowed to vary between $0.1<z<3$ in increments of 0.1. Namely, for each redshift, the galaxy templates were re-binned onto a grid with a resolution of 7.7\,\AA\ and rescaled to match the GMOS spectral region blue-ward of the \lya{} line ($<8924$\,\AA). The rescaled galaxy model was then subtracted from the combined GMOS and Flamingos-2 spectrum, and quasar templates were fitted to the resulting residual spectrum.
Although the observed spectral shape at $\sim8200$\,\AA\ could in principle be reproduced by a galaxy at $z\sim1.04$ exhibiting a pronounced 4000\,\AA\ break associated with an old stellar population, none of the tested galaxy templates at any redshift simultaneously reproduced both the emission blue-ward of the \lya{} line and the near-infrared spectral shape. An example of a fit obtained using an elliptical galaxy template at $z=1.04$ is shown in \autoref{fig:P170+20spec_galaxy}.

\bibliography{sample701}{}
\bibliographystyle{aasjournalv7}

\end{document}